\documentclass{iopart}
\usepackage{times}
\usepackage{iopams,amstext}
\usepackage{graphicx}
\usepackage{hyperref}
\usepackage[numbers,sort&compress]{natbib}
\usepackage{booktabs}
\hypersetup{colorlinks=true,linkcolor=blue,citecolor=blue,urlcolor=blue}

\newcommand{\binom}[2]{{{#1}\choose{#2}}}

\newcommand{\cL}{\mathcal{L}}

\newcommand{\const}{\mathop\mathrm{const}}
\newcommand{\nuMSM}{\ifmmode\nu\mathrm{MSM}\else$\nu$MSM\fi}
\newcommand{\dm}{\partial_\mu}

\renewcommand{\d}{\partial}
\newcommand\dd{\text{d}}

\begin{document}

\title{The Higgs field as an inflaton}
\author{Fedor Bezrukov}
\address{Physics Department, University of Connecticut, Storrs, CT
  06269-3046, USA,\\
  RIKEN-BNL Research Center, Brookhaven National Laboratory, Upton,
  NY 11973-5000, USA}
\ead{Fedor.Bezrukov@uconn.edu}

\begin{abstract}
  The Higgs field of the pure Standard Model can lead to the
  inflationary expansion of the early Universe if it is non-minimally
  coupled to gravity.  The model predicts Cosmic Microwave Background
  (CMB) parameters in perfect agreement with the current observations
  and has implications for the Higgs boson mass.  We review the model,
  its predictions, problems arising with its quantization and some
  closely related models.
\end{abstract}


\submitto{\CQG}
\noindent{\small\rm Preprint: RBRC1042\par}

\section{Introduction}

The current results from the Large Hadron Collider (LHC) completed the
particle zoo of the Standard Model (SM) of particle physics by the
discovery of a Higgs boson \cite{CMS:2012gu,ATLAS:2012gk}, while at
the same time showing no signs of any beyond the SM physics.  The
first results from the Planck satellite further confirm the
inflationary scenario for the early Universe \cite{Ade:2013uln} and
favour a simple inflationary scenario with only one slow rolling
scalar field.  The question remains, whether it is needed to add more
particle states to the SM to make the inflation possible, or can it be
explained with the already present ones?

Usual considerations require for the potential of the scalar field,
responsible for inflation, to be extremely flat.  The initial density
perturbations can be generated with the observed amplitude by a
chaotic inflation with a scalar field with the quartic self-coupling
$\lambda\sim10^{-13}$ or mass $m\sim10^{13}$\,GeV
\cite{Linde:1983gd}.  There are no particles in the SM with such
properties, but it turns out that there is another option.
Modification of the gravitational interactions of a scalar particle
\begin{equation}
  \label{dSNM}
  \delta S_{\text{NM}} =
  \int \dd^4x\sqrt{-g} \Big[ -\xi\Phi^\dagger\Phi R \Big] ,
\end{equation}
can lead to a good slow-roll behaviour for arbitrary large $\lambda$
(at the cost of large coupling $\xi$).  A field $\Phi$ with a large
(order $0.1$) quartic self-coupling is present in the SM---the Higgs
boson.  The inflationary model using the SM Higgs boson with the
non-minimal coupling (\ref{dSNM}) will be called Higgs inflation (HI).

The first step in the analysis of the inflation with the non-minimal coupling
is the classical solution of the slow-roll evolution.  It can be
simplified (as far as the equations of motion for the action with the
addition of (\ref{dSNM}) look rather complicated \cite{Kaiser:1994vs})
by a suitable change of the dynamical variables, which removes the
non-minimal coupling at the cost of the modification of the scalar
potential (and interactions with other particles of the theory).  This
variable change is traditionally called `conformal transformation',
and the theory in the new variables is referred to as the theory in
the `Einstein frame' (as opposed to the original `Jordan frame').
After this change all the formulas for the standard single-field
inflationary slow roll, and generation of the density perturbations can
be used, with the result being in extraordinary good agreement with
the Planck observations.  The details of the full model are described
in section~\ref{sec:nmcoupling}.

The inflationary scenario in HI depends on the absence of additional
new physics above the electroweak scale, but of course variations of
the setup are possible.  Some examples are described in
section~\ref{sec:variants}.  All other experimental observations,
required beyond the SM physics, namely neutrino oscillations, dark
matter and baryon asymmetry of the Universe can be explained with the
extensions of the SM that do not affect its properties above the
electroweak scale.  One possibility is an extension of the SM with
right-handed (sterile) neutrinos, for example, $\nu$MSM
\cite{Asaka:2005an,Asaka:2005pn,Boyarsky:2009ix}.

A significantly more complicated analysis is required beyond the
tree-level approximation.  With $\xi\gg1$, as required by the CMB
observations, the operator (\ref{dSNM}) introduces new scales in the
model below the Planck mass: $M_P/\xi$ and $M_P/\sqrt{\xi}$, and the
model requires some form of the high-energy (UV) completion.  While
the tree-level calculations of the inflation and generation of the
primordial density perturbations are performed at lower values of momenta,
the predictions of the model with the quantum corrections depend on
the completion.  First, the standard approach of perturbation theory, well
formulated for a theory above a trivial vacuum, cannot be applied,
and even in the analysis of perturbations on top of a classical background
(which is relevant for inflation), it is impossible to move beyond the
lowest order tree-level calculations without additional assumptions.
Various assumptions about the completion of the theory up to the high
momentum scales
allow us to make different levels of predictions.  The weaker one, scale
invariance for the theory in UV, allows us to work within a form of
effective theory during inflation.  A stronger set of assumptions,
namely the absence of the quadratic divergences, together with a definite
prescription for the subtraction rules for the divergent diagrams,
allows us to relate the low-energy and inflationary sectors of the
theory.  The role of the various scales is discussed in
section~\ref{sec:treeunitarity}, while the calculation of the quantum
corrections is outlined in section~\ref{sec:loops}.

As far as there are no additional scales in the model between the
electroweak and inflationary scales, a bound can be made on the Higgs
mass.  If the Higgs mass is low, the effective Higgs potential may
become negative at high scales.  If this happens below the
inflationary scale $M_P/\sqrt{\xi}$, the inflation will end up in this
new minimum of the potential, instead of the electroweak phase.  An
even more stringent bound may exist from the modification of the
inflationary potential by loop corrections, but this bound depends on
the details of the quantization of the theory.  The recent LHC results
gave the Higgs mass (within the experimental and theoretical errors)
at the boundary value (see section~\ref{sec:mass}).

\section{Non-minimal coupling to gravity and inflation}
\label{sec:nmcoupling}

\subsection{Inflationary scalar sector and inflation}
\label{sec:infsector}

One of the simplest inflationary setups is the chaotic inflation
\cite{Linde:1983gd}.  In this approach, the field starts at a large value
and rolls slowly towards its origin.  The slow roll is obtained due to
the `Hubble friction' connected with the expansion of the universe.
Then, if one requires the density perturbations of the field,
generated during inflation, to have the observed value, a significant
constraint on the potential is obtained.  For the quartic potential
this means, in particular, that the coupling constant is of the order
of $\lambda\sim10^{-13}$.  Such particle is not present in the SM, and
it is not easy to have this particle coupled to the SM, as far as for
not very small couplings they would induce radiative corrections,
spoiling the flatness of the potential.  However, there is a way out
of this unfortunate situation.  Let us note that a scalar field in
general may be coupled non-minimally to gravity, i.e.\ in addition to
the trace of the energy-momentum tensor gravity may `feel' the
scalar field background.  This type of coupling is even inevitable, as
renormalization of a scalar field in the curved spacetime requires
introduction of divergent counter-terms of this form
\cite{Birrell:1982ix}.  The action of the scalar field with this term
is
\begin{equation}
  \label{SJ}
  S_{J} = \int \dd^4x \sqrt{-g} \left[
    - \frac{M^2}{2}R - \xi\frac{h^2}{2}R
    + \frac{\partial_\mu h\partial^\mu h}{2}
    - V(h)
  \right] ,
\end{equation}
where $h$ denotes the scalar field, and the potential for the case of
the Higgs field is
\begin{equation}
  \label{V}
  V(h)=\frac{\lambda}{4}(h^2-v^2)^2 ,
\end{equation}
with the vacuum expectation value of the field $v$.  The metric
signature used is $(+,-,-,-)$, and the special value, corresponding to
the conformal coupling, is $\xi=-1/6$.  Note that in the case of the
Higgs field $h$ corresponds only to the radial mode,
$\Phi^\dagger\Phi=h^2/2$, or in the unitary gauge,
$\Phi=\frac{1}{\sqrt{2}}\binom{0}{h}$.  The effects from the angular
modes and interactions with other particles will be discussed later in
sections~\ref{sec:Sint} and \ref{sec:loops}.  The first two terms lead
to an effective change of the Planck mass in the presence of the
scalar background $M_{P,\text{eff}}^2\sim M^2+\xi h^2$.

It is rather cumbersome to analyse inflation directly with the action
(\ref{SJ}) (in the `Jordan frame').  The equations of motion are
entangled between the field and the metric (see
e.g.~\cite{Kaiser:1994vs}), the quadratic part of the action for the
small perturbations contains kinetic mixing between the perturbations
of the scalar field and the trace of the metric.  The simplest way to
work with this action is to get rid of the non-minimal coupling to
gravity by a suitable change of variables, which is conventionally
called conformal transformation from the Jordan frame to the Einstein
frame\footnote[1]{The equivalence of the analysis at the \emph{tree
    level} in both the Jordan and Einstein frames was studied by
  several authors \cite{Tsujikawa:2004my,Makino1991,Fakir:1992cg}.
  Actually, at the tree level the comparison is effectively reduced to
  the observation that with the one-to-one correspondence of the
  classical solutions in any variables, the perturbations on top of it
  in different frames (or, more precisely, different variable choices)
  are connected by a linear transformation and are equivalent.  At
  higher orders, the situation is more involved due to
  non-renormalizability of the theory, see section~\ref{sec:loops}.}
(see, e.g., \cite{Kaiser:1994vs,Tsujikawa:2000wc}):
\begin{equation}
  \label{gmunuOmega}
  g_{\mu\nu} \to \hat{g}_{\mu\nu} = \Omega^2 g_{\mu\nu} , \qquad
  \Omega^2 = \frac{M^2+\xi h^2}{M_P^2} ,
\end{equation}
where $M_P\equiv 1/\sqrt{8\pi G_N}=2.44\times10^{18}$\,GeV is the
reduced Planck mass.  The parameters $M$ and $M_P$ differ for the non-zero
vacuum expectation value of $\langle h\rangle=v$, but for most of the
situations, we will have $\xi v\ll M$ and set $M\simeq M_P$ (with
exceptions of sections \ref{sec:induced-gravity} and
\ref{sec:unitaryHI}, where the vacuum contribution of the
non-minimally coupled term actually dominates the Planck mass).  This
transformation leads to a non-minimal kinetic term for the Higgs
field.  It is also convenient to replace $h$ with a new canonically
normalized scalar field $\chi$, satisfying the relation
\begin{equation}
  \label{dchidh}
  \frac{\dd\chi}{\dd h} = \sqrt{
    \frac{\Omega^2+\frac{3}{2}M_P^2{(\Omega^2)'}^2}{\Omega^4} }
  = \sqrt{
    \frac{1+(\xi+6\xi^2)h^2/M_P^2}{(1+\xi h^2/M_P^2)^2} } ,
\end{equation}
where $'$ means the derivative over $h$.  Note that this redefinition is
only possible for one real scalar field, and the generic situation
with several scalar fields leads to the complicated structure of the
kinetic terms (see discussion in section~\ref{sec:Sint} and
\cite{Kaiser:2010ps}).
Finally, the action in the Einstein frame is
\begin{equation}
  \label{SE}
  S_E = \int \dd^4x\sqrt{-\hat{g}} \left[
    - \frac{M_P^2}{2}\hat{R}
    + \frac{\dm\chi\partial^\mu\chi}{2}
    - U(\chi)
  \right] ,
\end{equation}
where $\hat{R}$ is calculated using the metric $\hat{g}_{\mu\nu}$ and
the potential is rescaled with the conformal factor
\begin{equation}
  \label{U}
  U(\chi) = \frac{1}{\Omega^4\left[ h(\chi) \right]} 
            \frac{\lambda}{4}\left[ h^2(\chi)-v^2 \right]^2 .
\end{equation}
We will, rather ambiguously, write the potential $U$ and the scale factor 
$\Omega$ as the functions of either $h$ or $\chi$, which should not lead
to problems, as far as $h$ and $\chi$ can be expressed one through
another in a unique way with the solution of (\ref{dchidh}).

The analysis of the inflation in the Einstein frame is now
straightforward and follows the usual slow-roll approach.  The
slow-roll parameters (in notations of \cite{Linde:2007fr}) are easy to
express analytically as the functions of the field $h$ using
(\ref{dchidh}) and (\ref{U})
\begin{eqnarray}
  \label{genepsilon}
  \epsilon = \frac{M_P^2}{2}\left(\frac{\dd U/\dd\chi}{U}\right)^2
    = \frac{M_P^2}{2}\left(\frac{U'}{U}\frac{1}{\chi'}\right)^2, \\
  \label{geneta}
  \eta = M_P^2\frac{\dd^2U/\dd\chi^2}{U}
    = M_P^2\frac{U' \chi'-U' \chi'}{U {\chi'}^3} ,
\end{eqnarray}
with $'$ meaning derivative over $h$.  Slow-roll ends at
$\epsilon\simeq1$, with the corresponding field value $h_\text{end}$.
The observed modes left the horizon when the field value equals
$h_N$ which is determined by the number of inflationary $e$-foldings in
the Einstein frame,
\begin{equation}
  \label{N}
  N = \int\limits_{h_{\text{end}}}^{h_N}
    \frac{U \chi'^2}{U'}\frac{\dd h}{M_P^2}
  \simeq \frac{3}{4}\left[
    \left(\xi+\frac{1}{6}\right)\frac{h_N^2-h_{\text{end}}^2}{M_P^2}
    - \ln\frac{1+\frac{\xi h_N^2}{M_P^2}}
              {1+\frac{\xi h_\text{end}^2}{M_P^2}}
  \right] ,
\end{equation}
where we neglected the quadratic part of the potential.  Note that in
the Jordan frame the additional ratio of conformal factors enters in
the $e$-folding number leading to another definition
\cite{Lerner:2009na}.  The $N$ itself is determined from the scale of
the observed mode and evolution of the Universe after the inflation,
see section~\ref{sec:reheating}.  To generate the proper amplitude of the
density perturbations, the potential should satisfy at $h_N$ the
normalization condition
\begin{equation}
  \label{WMAPnorm}
  \frac{U}{\epsilon} 
  = 24\pi^2\Delta_\mathcal{R}^2M_P^4\simeq(0.0276M_P)^4 .
\end{equation}
With the non-minimal coupling as an additional parameter this
condition becomes a relation between $\xi$ and $\lambda$, instead of a
condition on just $\lambda$.  The inflationary predictions (see, e.g.,
\cite{Linde:2007fr}) for the CMB spectrum parameters are then given by
the expressions for the spectral index $n_s$ and the tensor-to-scalar
perturbation ratio $r$,
\begin{equation}
  \label{gen-nsr}
  n_s = 1-6\epsilon+2\eta ,\qquad
  r = 16\epsilon ,
\end{equation}
calculated at $h_N$.

Let us immediately note a couple of the features of the obtained action.
First, if the Jordan frame potential $V(h)$ is quartic, then, for high
values of the field $\xi h^2\gg M_P^2$, the resulting Einstein frame
potential $U$ is flat.  This suggests that the resulting theory
provides a `better' inflation than the minimally coupled one.  See
\cite{Park:2008hz,Tsujikawa:2004my,Tsujikawa:2000wc} for the discussion of
other power-law potentials and non-minimal couplings.

\subsection{Interaction with other particles}
\label{sec:Sint}

Let us describe the interaction of the scalar with other particles,
neglected in the previous section.  First, note that the Higgs boson
itself is a complex doublet $\Phi$.  The generalization of the
conformal transformation (\ref{gmunuOmega}) is rather obvious: using
$\Omega^2=1+2\xi\Phi^\dagger\Phi/M_P^2$ as the conformal factor, we
obtain the Einstein frame action as
\begin{equation}
  \label{SEFdoublet}
\fl  S_\text{E} = \int \dd^4x\sqrt{-g} \Bigg[
    - \frac{M_P}{2}\hat{R}
    + \frac{3M_P^2}{4}
      \frac{\partial_\mu\Omega^2}{\Omega^2}
      \frac{\partial^\mu\Omega^2}{\Omega^2}
   + \frac{D_\mu\Phi^\dagger D^\mu\Phi}{\Omega^2}
    - \frac{V(\Phi^\dagger\Phi)}{\Omega^4}
  \Bigg] ,
\end{equation}
where $D_\mu$ is the gauge covariant derivative.  Note that only the
kinetic term for the radial mode can be taken into the canonic form by
the field redefinition (\ref{dchidh}), while the derivative
interaction of the other components with the radial mode (third term)
cannot be removed in general.  The gauge boson and fermion kinetic
terms are invariant (at least classically) under the conformal
transformations (\ref{gmunuOmega}) augmented by the field
transformations
\begin{equation}
  \label{Amupsichange}
  A_\mu \to A_\mu ,\qquad
  \psi \to \hat\psi = \Omega^{-3/2}\psi .
\end{equation}
The only terms that are changed are the massive terms.  For the gauge
bosons, this is due to the rescaling of the third term in
(\ref{SEFdoublet}), and for the fermions, the Yukawa terms change to
\begin{equation}
  \label{EFyukawa}
  S_{E,\text{Yukawa}} = \int \dd^4x\sqrt{-\hat{g}} \left[
    y \bar\psi_R \frac{\Phi^\dagger}{\Omega} \psi_L + \cdots
  \right] .
\end{equation}
In the unitary gauge $\Phi=\frac{1}{\sqrt{2}}\binom{0}{h}$, these
changes mean the change $h \to h(\chi)/\Omega(\chi)$ in all the gauge
boson and fermion mass terms.

As in the potential, the powers of $h$ and $\Omega$ are the same in
the numerator and denominator, so that the Einstein canonically
normalized field $\chi$ always enters the action in the same
combination.  Specifically, during inflation (at high values of the
field), the radial scalar mode decouples from the other fields, while
leaving all particle masses proportional to $M_P/\sqrt{\xi}$.

If there are additional scalars, the modifications also involve the
modifications in kinetic terms, analogously to the third term in
(\ref{SEFdoublet}) for a minimally coupled scalar (see
e.g.~\cite{Kaiser:2010ps,Kaiser:2012ak,Kaiser:2013sna}).

A convenient description of the SM in the Einstein frame can be
obtained by the definition
$\Phi=\exp[2i\pi(x)^aT^a]\binom{0}{h(x)/\sqrt{2}}$, with the SU(2)
generators $T^a=\tau^a/2$.  Then the action for the scalar mode
$h(\chi)$ is (\ref{SE}), and the rest corresponds to the chiral
electroweak action with the `vacuum expectation value' set to
$h(\chi)/\Omega$, see \cite{Bezrukov:2009db,Dutta:2007st}.

\section{Predictions for cosmology, universe history}
\label{sec:cosmology}

\subsection{Inflationary stage}
\label{sec:inflationary-stage}
 
There are several interesting limiting situations, depending on the
choice of the parameters in the potential and gravitational
sector \cite{Salopek:1988qh,Kaiser:1994vs}.

\subsubsection{Large $\xi$ limit (Higgs inflation)}

\begin{figure}
  \centering
  \includegraphics[width=0.5\textwidth]{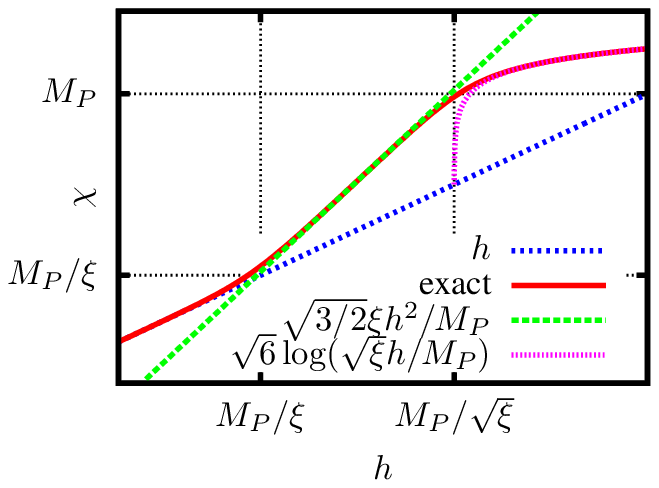}%
  \raisebox{5ex}{\hbox{\includegraphics[width=0.5\textwidth]{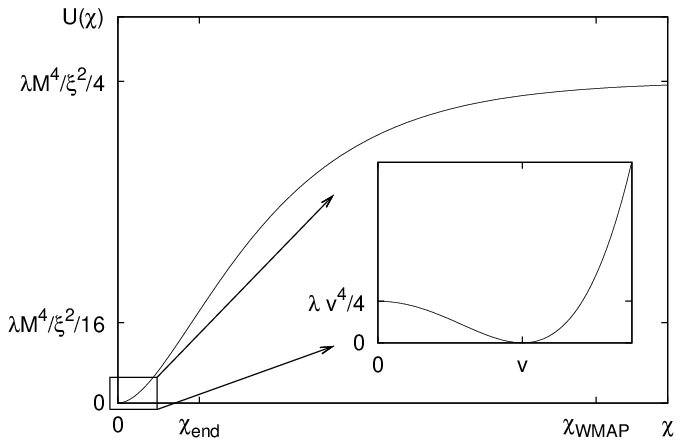}}}
  \caption{Left: dependence of $\chi$ (the Einstein frame Higgs field)
    on $h$ (the Jordan frame Higgs field), logarithmic scale.  Right:
    effective potential in the Einstein frame.  The insert
    magnification is not to scale.}
  \label{fig:chihUeff}
\end{figure}

The HI case corresponds to the large $\xi\gg1$ and negligible vacuum
contribution of the non-minimal coupling to the Planck mass, $\xi v\ll
M_P$.  Then equation (\ref{dchidh}) can be solved approximately as
\begin{equation}
  \label{chi(h)}
  \chi \simeq \cases{
    h & for $h \ll \frac{M_P}{\xi}$ ,\\
    \textstyle \sqrt{\frac{3}{2}}M_P\ln\Omega^2(h) & for $h\gg\frac{M_P}{\xi}$
    .\\
  }
\end{equation}
This relation is illustrated in figure~\ref{fig:chihUeff}.  Using this
relation, we can explicitly write the potential for the large Higgs
background $h,\chi\gg M_P/\xi$ as
\begin{equation}
  \label{U(chi)}
  U(\chi) \simeq \frac{\lambda M_P^4}{4\xi^2} \left(
    1 - \e^{-\frac{2\chi}{\sqrt{6}M_P}}
  \right)^2 .
\end{equation}
For a small field $h,\chi\ll M_P/\xi$, the potential turns into the usual
SM quartic potential with a negative mass.  Figure~\ref{fig:chihUeff}
schematically shows the potential.

Note that there are actually two interesting energy scales for
$\xi\gg1$.  The lowest one is $M_P/\xi$, corresponding to the scale
where deviation from the SM becomes significant.  The higher one is
$M_P/\sqrt{\xi}$, which is the energy scale at inflation and typical
particle scale for excitation masses during inflation.  The Hubble
scale (in the Einstein frame) during inflation is $H^2\simeq\lambda
M_P^2/(12\xi^2)$.

The potential (\ref{U(chi)}) is exponentially flat for large field
values and provides the slow-roll inflation.  From
(\ref{genepsilon}), (\ref{geneta}) and (\ref{V}), we obtain
\begin{equation}
  \epsilon \simeq \frac{4 M_P^4 }{3\xi^2h^4} ,\qquad
  \eta \simeq
  \frac{4 M_P^4}{3 \xi^2 h^4 }\left(1-\frac{\xi h^2}{M_P^2}\right) .
\end{equation}
Inflation ends at
$h_\text{end}\simeq(4/3)^{1/4}M_P/\sqrt{\xi}\simeq1.07M_P/\sqrt{\xi}$
(and $\chi_\text{end}\simeq 0.94M_P$).  Using (\ref{N}), we obtain the
field value corresponding to the horizon crossing of the observed CMB
modes $h_{N}\simeq9.14M_P/\sqrt{\xi}$.  Thus, the CMB normalization
(\ref{WMAPnorm}) requires
\begin{equation}
  \label{xi-numeric}
  \xi \simeq 47\,000\sqrt{\lambda} ,
\end{equation}
where $\lambda$ is the Higgs boson self-coupling constant, taken at
an inflationary scale.

\begin{figure}
  \centering
  \includegraphics[width=\textwidth]{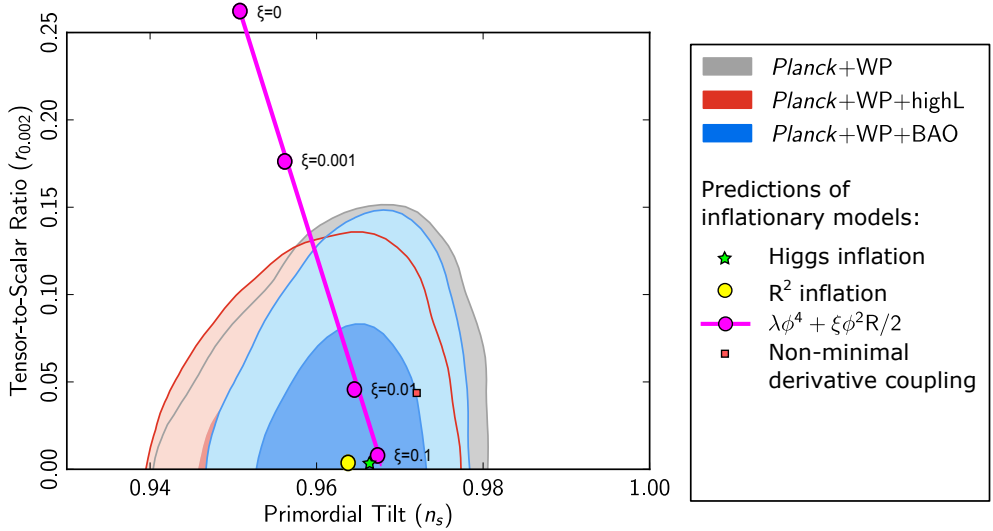}
  \caption{Predictions (at a tree level, which coincides with the
    radiatively corrected results in the scale-invariant quantization,
    choice I of section \ref{sec:mass}) from the inflationary models
    \cite{Bezrukov:2007ep,Bezrukov:2011gp,Bezrukov:2013fca} and the
    Planck satellite \cite{Ade:2013uln} observed bounds.}
  \label{fig:WMAP}
\end{figure}

Finally, to the lowest order in $1/\xi$, the spectral index and the
tensor-to-scalar perturbation ratio are
\begin{equation}
  \label{ns-r}
  n_s \simeq 1-8\frac{4N+9}{(4N+3)^2} \simeq 0.967 ,\qquad
  r \simeq \frac{192}{(4N+3)^2} \simeq 0.0031 ,
\end{equation}
where the number of $e$-foldings $N\simeq57.7$ (see section
\ref{sec:reheating}).  Note that there are higher corrections
suppressed by $1/\xi$ and $1/N$ to these formulae.  See
figure~\ref{fig:WMAP} for the good agreement of this result with the
latest measurements.  The inflation model in the Einstein frame is a
simple one field slow-rolling inflation, with all extra degrees of
freedom much heavier than the Hubble scale ($m\sim M_P/\sqrt{\xi}\gg
H\sim M_P/\xi$), so it does not predict any significant
non-Gaussianities in the spectrum.

\subsubsection{Nearly quartic inflation}

Let us note that the case of small $\xi$, though not relevant for HI,
provides a variation of the usual quartic potential inflation which is
in agreement with the recent experiments.  The analysis for this case
can be performed by approximating (\ref{N}) as (which by coincidence
works quite well for all values of $\xi$)
\begin{equation}
  \label{interpolating-field}
  (1+6\xi) \frac{h_N^2}{M_P^2} \simeq 8(N+1) .
\end{equation}
Plugging this into (\ref{genepsilon}), (\ref{geneta}) and
(\ref{gen-nsr}), the CMB predictions are obtained, see
\cite{Bezrukov:2013fca} for details.  The result smoothly interpolates
with the growing $\xi$ between the minimally coupled (now strongly
disfavored by Planck) and HI points.  For $\xi\gtrsim0.003$ the
predictions enter into the $2\sigma$ experimentally allowed region, without
introducing any new scales of the type discussed in sections
\ref{sec:treeunitarity} and \ref{sec:loops}.  The required quartic
coupling, however, is still very small, nearly unchanged from the pure
quartic value $\lambda\simeq 1.5\times10^{-13}$, so a dedicated
inflaton field is required.

\subsubsection{Induced gravity}
\label{sec:induced-gravity}

Another interesting case emerges when the Planck mass is generated by the
vacuum expectation value of the scalar field via the non-minimal
coupling term, $M_P^2=\xi v^2$.  This corresponds to the very early
works about the non-minimal coupling
\cite{Zee:1978wi,Smolin:1979uz,Spokoiny:1984bd,Fakir:1990eg,%
Salopek:1988qh,Cooper:1982du,CervantesCota:1994zf,CervantesCota:1995tz}.
With the non-minimal coupling given by (\ref{xi-numeric}), the value
for $v=M_P/\sqrt{\xi}$ is relatively close to the GUT scale for
$\lambda\sim1$.  This suggests that the identification of the inflaton
field may be made with the GUT scale Higgs boson. The conformal factor
is in this case
\begin{equation}
  \label{Omega-induced}
  \Omega^2 = \frac{\xi h^2}{M_P^2} \equiv \frac{h^2}{v^2} .
\end{equation}
The relation between the Jordan and Einstein frame fields is obtained
in an exact form from (\ref{dchidh}) and (\ref{Omega-induced}) as
\begin{equation}
  \label{h-chi-induced}
  h = v \exp\left( \frac{\chi}{\sqrt{6+1/\xi}M_P} \right) ,
\end{equation}
The Einstein frame potential is now equal to (\ref{U(chi)}) for all
values of the field.
However, the field $\chi$ becomes \emph{exactly} decoupled from all
the gauge (and fermion) fields it was giving mass to.  As described in
section~\ref{sec:Sint} all the masses of the gauge bosons and
fermions, associated with the Higgs boson, are proportional to
$h/\Omega$. With the conformal factor (\ref{Omega-induced}), this is
constant $M_P/\sqrt{\xi}$.  Thus, the scalar field completely
decouples from the matter fields and couples to them only via
gravity.  Note, however, that it \emph{is} coupled to all other
non-conformal terms in the actions, specifically to the other scalar
particles (e.g.\ the SM Higgs boson) via the conformal factor.  These
interactions are suppressed by the Planck mass in the exponent of
(\ref{h-chi-induced}).

A singlet scalar of this type is an exact equivalent of the scalaron
which appears after the conformal transformation of $R^2$ gravity
\cite{Starobinsky:1980te}, and its inflationary phenomenology was
described recently in detail in
\cite{Gorbunov:2010bn,Bezrukov:2011gp,Gorbunov:2012ij,Gorbunov:2012ns}.

\subsection{Post inflationary cosmology}
\label{sec:reheating}

Another notable feature of the HI model is that the evolution of the
Universe after inflation, i.e.\ a preheating mechanism is fixed.  This
removes the ambiguity in the CMB parameter definition associated with
the unknown number of $e$-foldings of inflation, which is often
indicated on the plots with inflationary predictions.

The preheating in the model was analysed in
\cite{Bezrukov:2008ut,GarciaBellido:2008ab}.  After the end of
inflation, $\chi<M_P$ ($h<M_P/\sqrt{\xi}$), the potential
(\ref{U(chi)}) can be approximated by a quadratic function (with just
a small quartic part around the origin corresponding to the SM
potential for $\chi<M_P/\xi$).  The field $\chi$ starts to oscillate
and the expansion of the universe corresponds to the matter dominated
regime.  During this regime the SM particles (mostly $W$ and $Z$
bosons) are produced at the moments when the Higgs field crosses zero.
At first, they decay faster than the oscillations of the field $\chi$,
not leading to effective preheating (transfer of energy from the
inflaton to the relativistic SM degrees of freedom).  At the moment
when they start to live long enough the parametric resonance starts,
the concentration of the gauge boson rises and they start to
effectively annihilate into light SM particles \cite{Bezrukov:2008ut},
leading to nearly immediate preheating.  Another significant reheating
mechanism is the production of relativistic excitations of the Higgs
field itself, which leads to the lower bound on the reheating
temperature.  The preheating temperature is estimated as $T_r\sim
0.3\text{--}1.1\times10^{14}$\,GeV.  Numerically more precise analysis
of \cite{GarciaBellido:2008ab} leads to similar results for the onset
of parametric resonance.  There are a number of effects which are not
yet carefully taken into account in the reheating analysis (Pauly
blocking of the produced fermions, the decay of the gauge bosons at
the moments of the zero crossing, when the particle description fails,
back reaction effects, etc), so the value of the reheating
temperature may still differ a bit from the given values.

Note that although determining the temperature when the whole matter in
the universe is properly thermalized may be a hard problem, it is easy
to estimate the absolute lower bound on the effective temperature when
the expansion of the universe switches to the radiation dominated
regime (which is the moment relevant for the calculation of the number
of $e$-foldings in the next paragraph).  For this, let us note that at
$h\lesssim M_P/\xi$ the Higgs potential switches to the quartic one,
meaning that the expansion is effectively radiation dominated.
Equating the potential to the thermal energy density at this moment one
obtains the estimate of
\begin{equation}
  T_r \gtrsim \left(\frac{15}{2\pi^2g_*\lambda}\right)^{1/4}
              \frac{M_p}{47\,000}
      \gtrsim 10^{13}\,\mathrm{GeV},
\end{equation}
so further calculation of the reheating is not going to change the
$e$-folding number beyond the error quoted in (\ref{Nreheating}).  Note
that this bound moves up for smaller Higgs masses.

The number of $e$-foldings in the Einstein frame (where the standard
sigle-field inflationary analysis can be used) between the modes
corresponding to the present day pivot scale $k/a_0$ exiting the
horizon and the end of inflation is
\begin{eqnarray}
  \label{Nreheating}
  N &= \ln\left(\frac{\sqrt{\pi}}{270^{1/4}}
           \frac{\tilde{g}_0^{1/3}}{g_{*r}^{1/12}}\right)
  -\ln\frac{k}{a_0T_0}
  - \ln\frac{M_P}{U_N^{1/4}} \frac{U_e^{1/4}}{U_N^{1/4}}
  - \frac{1}{3}\ln\frac{U_e^{1/4}}{\rho_r^{1/4}} \nonumber\\
  &= 58.9-\frac{1}{3}\ln\frac{U_e^{1/4}}{\rho_r^{1/4}}
  \simeq 57.7\pm0.2 ,
\end{eqnarray}
where $k/a_0=0.002$\,Mpc is the scale of the observed perturbations
used for the normalization of the spectrum, $g_{*r}=106.75$ and
$\tilde{g}_0=43/11$ are the effective number of d.o.f. at reheating
and now, $T_0=2.725$\,K is the CMB temperature, $U_N=U(h_N)$ and
$U_e=U(h_\text{end})$ are the Einstein potentials at the moment of
horizon crossing and at the end of slow roll and $\rho_r$ is the thermal
energy density at preheating.  The error here comes from the reheating
estimates in \cite{Bezrukov:2008ut} and may be somewhat underestimated.

Knowing the exact value of $N$ allows for the precise prediction of the
spectral index and tensor-to-scalar ratio (\ref{ns-r}).  In fact, this
precise prediction even allows us to distinguish between the HI and the
$R^2$ inflation, both of which have similar inflationary potentials, but
different preheating mechanisms \cite{Bezrukov:2011gp,Gorbunov:2012ns}.


\section{Naturalness and strong coupling---tree-level analysis}
\label{sec:treeunitarity}

Let us find the scale associated with the non-minimal coupling
(\ref{dSNM}).  To do this we expand the metric on top of the flat
background $\eta_{\mu\nu}$ (we work here in the Jordan frame, the
Einstein frame result is the same, but is slightly more cumbersome to
obtain, see \cite{Bezrukov:2010jz})
\begin{equation}
  \label{gmnuhmunu}
  g_{\mu\nu} = \eta_{\mu\nu}+\frac{1}{M_P}\gamma_{\mu\nu} ,
\end{equation}
where $\gamma_{\mu\nu}$ has a canonically normalized kinetic term.
The leading interaction term from (\ref{dSNM}) is
\begin{equation}
  \label{xiRflatexp}
  \frac{\xi}{M_P}h^2\eta^{\mu\nu}\partial^2 \gamma_{\mu\nu} .
\end{equation}
On dimensional grounds this term makes the tree-level scattering
amplitude grow at high energies, and eventually violates the
unitarity bound \cite{Cornwall:1974km} at the energies of the order of $E\sim
M_P/\xi$
\cite{Burgess:2009ea,Barbon:2009ya,Burgess:2010zq,Hertzberg:2010dc,Atkins:2010yg}
(there are some cancellations for the case of a real field $h$ without
a potential, which are no longer true for complex or interacting scalar
fields \cite{Hertzberg:2010dc}).  This scale is low; for example, it is
much smaller than the energy density scale at inflation
$\lambda^{1/4}M_P/\sqrt{\xi}$ or inflationary field values $h\sim
M_P/\sqrt{\xi}$.  It is still slightly (parametrically in
$\sqrt{\lambda}$) larger than the inflationary Hubble scale (which is
the relevant momentum scale for processes during inflation)
$H\sim\lambda^{1/2}M_P/\xi$, but this is quite a tight call.

Following the standard rules of quantum field theory, this observation
is very dangerous for the Higgs inflation.  Actually, usual
assumptions of the effective field theories suggest that if there is
an energy scale $\Lambda\sim M_P/\xi$ in the theory, corresponding to
the violation of the tree-level unitarity, then some new physics
enters at this scale, and leads to the appearance of the operators
$h^{(4+n)}/\Lambda^n$ in the action, with some unknown coefficients.
It is then usually assumed that all these coefficients are some
uncorrelated numbers of order 1.  Thus, the analysis of the theory for
$h>\Lambda\sim M_P/\xi$ is assumed to be impossible, rendering the
inflationary analysis of the previous section (with $h\sim
M_P/\sqrt{\xi}\gg M_P/\xi$) false.  However, it is still possible to
obtain useful predictions from the HI, if one uses a more generic
understanding of effective field theories.  First, we will check that
tree-level calculations are safe if the classical background is taken
into account (instead of the vacuum one).  Then, we will formulate
assumptions about the theory that make the possible calculation of the
quantum corrections.  Weaker assumptions allow us to work in a form of
an effective theory in an inflationary regime and the low-energy
regime separately, and stronger assumptions make the connection
between these regimes possible.

Let us first study whether our lowest order approximation for the
universe evolution can be trusted (i.e.\ the inflation itself and the
generation of density perturbations proceed as described).  The
processes we study during inflationary expansion occur around the
classical background, that is, the slow-roll solution during the
inflation, so the (quantum) analysis should be made for the
perturbations $\delta\Theta(x)$ \emph{on top of the classical
  background $\Theta_0$} \cite{Bezrukov:2010jz,Ferrara:2010in}:
\begin{equation}
  \label{dTheta}
  \Theta(x) = \Theta_0 + \delta\Theta(x) ,
\end{equation}
where $\Theta$ stands for all the fields of the model.  To
estimate this cut-off scale $\Lambda(\Theta)$, we should substitute
(\ref{dTheta}) into the action and look for the expansion in powers of
$\delta\Theta$:
\begin{equation}
  \label{dh^n}
  \frac{\delta\Theta^n}{\Lambda_N(\Theta_0)^{n-4}},
\end{equation}
and take the minimal scale
$\Lambda(\Theta_0)=\min_N\Lambda_N(\Theta_0)$ (this rounds off some
corners, like possible cancellations or amplitude behaviour at large
$N$, which are irrelevant for the current discussion).  So, instead of
the same cut-off for the momentum and for the values of the field, we
have the energy scale depending on field background,
\emph{background-dependent cut-off scale}, below which the tree-level
calculation is possible\footnote{We often use the widely spread term
  `cut-off', while strictly speaking we always mean `the momentum
  scale of the violation of the tree-level unitarity'.}.  Note that
even with this definition the theory is not complete
\cite{Bezrukov:2010jz} and enters into a strongly coupled regime above
the energy $\Lambda(\Theta_0)$, where it is unitarized by some yet
unknown mechanism (own strong dynamics, asymptotic safety or even new
perturbative physics), cf \cite{Atkins:2010yg,Hertzberg:2011rc}.


The proposed approach can be carried both in the Jordan and Einstein
frames \cite{Bezrukov:2010jz}, bearing in mind that the cut-off scales
are connected by the conformal scaling
\begin{equation}
  \label{cutoffConformal}
  \Lambda_J = \Omega \Lambda_E .
\end{equation}
Here, we will review only the Jordan frame estimate; see
\cite{Bezrukov:2010jz} for details and the calculation in the Einstein
frame.  For the scalar-gravity sector, we have the expansion 
\begin{equation}
  \label{gphiexp}
  g_{\mu\nu} = \bar g_{\mu\nu}+\frac{1}{M_P}\gamma_{\mu\nu} ,\qquad
  h = \bar{h}+\delta h .
\end{equation}
The quadratic Lagrangian for the excitations has the form
\begin{eqnarray}
  \label{Lagr2}
  \cL^{(2)} = &
  - \frac{M_P^2+\xi\bar{h}^2}{8M_P^2} \left(
    \gamma^{\mu\nu}\Box\gamma_{\mu\nu}
    + 2\d_\nu\gamma^{\mu\nu}\d^\rho\gamma_{\mu\rho}
    - 2\d_\nu\gamma^{\mu\nu}\d_\mu\gamma
    - \gamma\Box\gamma
  \right) \nonumber\\
  &
  + \frac{1}{2}(\d_\mu\delta h)^2
  + \xi\bar{h}\left(
    \Box\gamma-\d_\lambda\d_\rho\gamma^{\lambda\rho}
  \right)\delta h ,
\end{eqnarray}
where $\gamma=\gamma^\mu_\mu$.  We retained here only the terms with
two derivatives of the excitations as they determine the UV behaviour
of the scattering amplitudes and hence the unitarity violation scale.
Note that in the non-trivial background, a large kinetic mixing
between the trace of the metric and the scalar perturbations is
present
\cite{Barvinsky:2008ia,DeSimone:2008ei,Barvinsky:2009fy,Barvinsky:2009ii}.
The change of variables
\begin{eqnarray}
  \label{chisubs}
  \delta h =
  \sqrt{\frac{M^2_P+\xi\bar{h}^2}{M^2_P+(\xi+6\xi^2)\bar{h}^2}}
  \,\delta\hat{h} ,\\
  \label{hsubs}
  \gamma_{\mu\nu} =
  \frac{M_P}{\sqrt{M^2_P+\xi\bar{h}^2}}\,\hat\gamma_{\mu\nu}
  - \frac{2\xi\bar{h}\bar{g}_{\mu\nu}}{\sqrt{(M^2_P+\xi\bar{h}^2)(M^2_P+(\xi+
      6\xi^2)\bar{h}^2)}}\delta\hat{h}
\end{eqnarray}
diagonalizes the kinetic term.  The unitarity violation scale can be
read out of the operators with a dimension higher than 4.  The
leading operator is the cubic scalar--graviton interaction $\xi(\delta
h)^2\Box\gamma$, which is written in terms of the canonical fields
$\delta\hat{h}$, $\hat\gamma_{\mu\nu}$ obtains the dimensional
coefficient corresponding to the cut-off scale
\begin{equation}
  \label{cutoff}
  \Lambda^J _\text{g-s}(\bar{h}) \sim \frac{M^2_P+(\xi+6\xi^2)\bar{h}^2}
                            {\xi\sqrt{M^2_P+\xi\bar{h}^2}} .
\end{equation}
At the low-field background $\bar{h}\ll M_P/\xi$, this formula readily
reproduces the $M_P/\xi$ cut-off of the works
\cite{Burgess:2009ea,Barbon:2009ya,Burgess:2010zq,Hertzberg:2010dc}.
It is smaller than the Planck mass, but for $\xi\sim10^4$, it is still
way above the reach of the collider experiments.

At large inflationary fields, $\bar{h}\gg M_P/\sqrt{\xi}$, the cut-off
grows linearly with $\bar{h}$ (or reaches Planck scale in the Einstein
frame), making the calculation of the inflationary perturbations safe.
Note that this coincides with the cut-off in the gravitational (spin-2)
sector.  The latter is given by the effective Planck mass defined as
the coefficient in front of the $R$ term in the Lagrangian,
$\Lambda^J_\text{Planck}=\sqrt{M^2_P+\xi\bar\phi^2}$.

\begin{figure}
  \centering
  \includegraphics[width=0.5\textwidth]{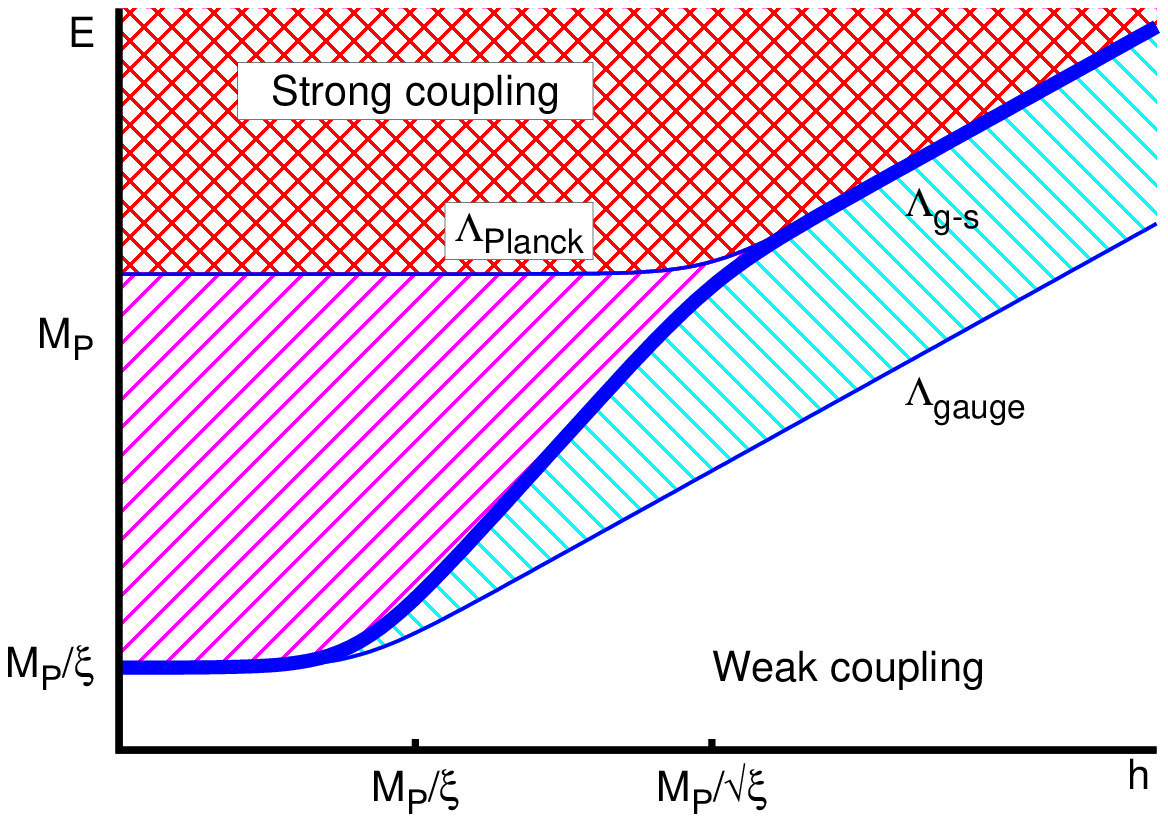}%
  \includegraphics[width=0.5\textwidth]{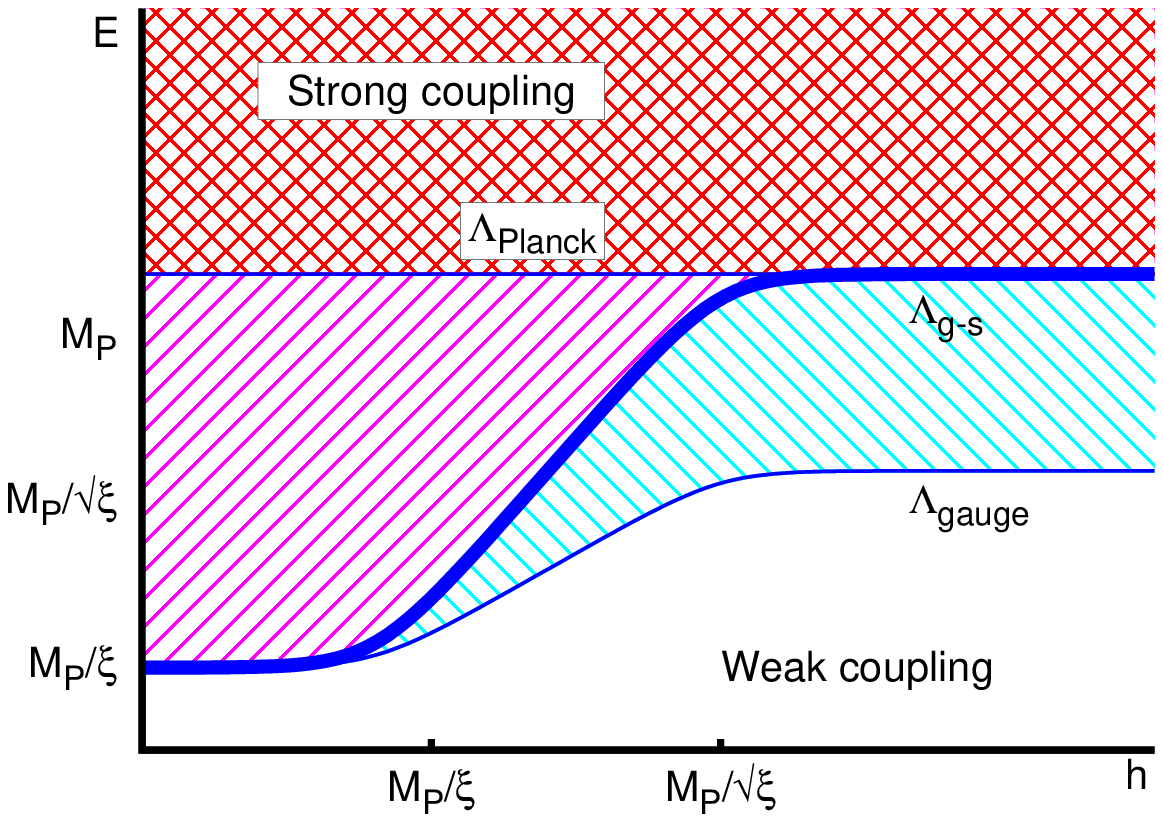}
  \caption{Schematic depiction of the cut-offs (scales of violation of
    the tree-level unitarity) in the Jordan (left) and Einstein
    (right) frames for different sectors of the theory.
    $\Lambda_\text{g-s}$ corresponds to the inflaton (scalar sector)
    scatterings, $\Lambda_\text{gauge}$ to the gauge boson scattering
    and $\Lambda_\text{Planck}$ is the graviton scattering.}
  \label{fig:cutoffs}
\end{figure}

In an analogous way one can estimate the cut-off in the gauge sector
(see \cite{Bezrukov:2012hx} for the detailed calculation in a closely
related Higgs-Dilaton model)
\begin{equation}
  \label{LambdaJgauge}
  \Lambda^J_\text{gauge}(\bar{h}) \sim
  \frac{\sqrt{M_P^2+\xi(1+6\xi)h^2}}{\sqrt{6}\xi} .
\end{equation}

The behaviour of the cut-offs (\ref{cutoff}) and (\ref{LambdaJgauge}) is
illustrated in figure~\ref{fig:cutoffs} both in the Jordan and Einstein
frames.

Thus, we can conclude that at a tree level the situation is safe.  At
inflation, the tree-level cut-offs (\ref{cutoff}) and (\ref{LambdaJgauge})
are significantly higher than the typical interesting momenta, which
are of the order of the Hubble scale $H\sim M_P/\xi$.  Currently,
(after preheating, $h<M_P/\xi$), we have small temperatures,
$T<M_P/\xi$, much below the cut-off.  The most subtle situation is
achieved during the preheating, especially at its early stages.
However, although it makes the careful preheating calculation even more
complicated, it is unlikely for us to expect that the final preheating
results will be significantly different, and the numerical impact of
the preheating on the CMB observations is encoded in the number of
$e$-foldings (\ref{Nreheating}) and is logarithmically weak.

One should note that as far as above some of the scales $\Lambda$
discussed in this section, new physics can be expected, additional
higher order operators can be present in the theory, suppressed by
this scale.  Various effects of these operators were analysed in
\cite{Bezrukov:2008ut,Bezrukov:2011sz}, including the possible generation
of the baryon asymmetry and sterile neutrino dark matter.

\section{Quantization and loop corrections}
\label{sec:loops}

Let us now move to the analysis of the loop corrections to the Higgs
inflation.  The question of how to calculate them really does not have
a unique answer, as far as the action (\ref{SJ}) is
non-renormalizable.  The non-renormalizability of the action means
that, in principle, an infinite number of counter-terms should be fixed
to obtain predictions from the theory.  Strictly speaking, performing
scattering experiments at energies below $M_P/\xi$ around the
electroweak vacuum background does not allow us to distinguish between
different potentials at inflation.  In the Einstein frame (for the
simplicity of the argument), the arbitrarily precise scattering
measurement gives a finite number of coefficients in the power series
for the Higgs potential around zero (electroweak vacuum).  Altering
the remaining infinite number of coefficients, it is possible to
imitate an arbitrary potential at high values of the field.  This
means that additional principles (or assumptions about the UV
completion) should be used to make the theory predictable.  This is
inevitable, as far as, though we are working within the boundaries of
allowed momenta, we are interested in large field backgrounds, and
that takes us beyond the limits of ordinary effective field theory.

The effective field theory around the electroweak background is easy
to describe: it is given by the usual SM with power-law corrections
suppressed by the $M_P/\xi$ scale.  To make more precise predictions
one needs to measure more coefficients.

A less ordinary setup is needed to obtain the effective field theory
at inflation.  The crucial observation is that at large Higgs field
values, the theory possesses an approximate symmetry---scale
transformations in the Jordan frame, or shifts of the Higgs field
$\chi\to\chi+\const$ in the Einstein frame.  In the Einstein frame, we
have the minimally coupled scalar field theory with the potential of
the form
\begin{equation}
  U(\chi)=U_0 \left(
    1+\sum_{n=1}^{\infty}u_n\e^{-\frac{2n\chi}{\sqrt{6}M}}
  \right) .
\end{equation}
Splitting the field into the perturbations and smooth background as
$\chi=\bar\chi+\delta\chi$, one can check \cite{Bezrukov:2010jz} that
all the divergent terms are proportional to the exponents (vanishing
at large $\chi$) and that all the divergences can be absorbed into
the Lagrangian of the form
\begin{equation}
  \label{scalagr1}
  \cL = f^{(1)}(\chi)\frac{(\d_\mu\chi)^2}{2}-U(\chi)
  +f^{(2)}(\chi)\frac{(\d^2\chi)^2}{M^2}
  +f^{(3)}(\chi)\frac{(\d\chi)^4}{M^4}+\cdots ,
\end{equation}
with the coefficient functions
\begin{equation}
  \label{genser}
  f^{(i)}(\chi) = \sum_{n=0}^\infty f_n^{(i)}\e^{-\frac{2n\chi}{\sqrt{6}M}} .
\end{equation}
Although an infinite number of terms are present in this Lagrangian,
inflationary predictions at any chosen precision need only a
\emph{finite} set of terms, as far as the exponents are small during
inflation and higher order terms are less relevant.

The same form of the contributions appears from the other SM fields,
because they couple to the Einstein frame Higgs field in the
combination $\e^{-2\chi/\sqrt{6}M_P}$, see section~\ref{sec:Sint}.
The asymptotic shift invariance of the whole action in the Einstein
frame (which originates form the scale invariance of the Jordan frame
action (\ref{SJ})) makes it possible to develop an effective
inflationary theory with self-induced radiative corrections under
control (cf~\cite{Cheung:2007st}).

The additional assumption used here is that no contributions, breaking
the shift invariance, appear from the UV completion of the theory.
Such corrections do not immediately appear from gravity, as the field
$\chi$ is coupled to gravity minimally in the Einstein frame
(non-perturbative gravitational effects may spoil the picture, but
they are model dependent and can be exponentially suppressed
\cite{Kallosh:1995hi}).  Scale invariance in the Jordan frame suggests
that the use of scale-invariant models may be important to address the
problem (see \cite{Shaposhnikov:2008xi,Bezrukov:2012hx}), but no full
scale-invariant UV completion is known at the moment.

Even more difficult is the relation of the physics around EW vacuum
and the effective inflationary theory.  To see this, let us try to
estimate the loop corrections to the potential, supposing that at the
tree level an exact potential $\lambda U(\chi)$ is known (we
explicitly show the coupling constant here).  The expansion on top of
the background $\bar\chi$ has the form
\begin{equation}
  \lambda U(\bar\chi+\delta\chi)=\lambda\left[U(\bar\chi)+
    \frac{1}{2}U'(\bar\chi)(\delta\chi)^2
    +\frac{1}{3!}U''(\bar\chi)(\delta\chi)^3+\cdots\right] .
\end{equation}
Computing loop corrections in, say, the cut-off
regularization scheme one generates the divergences of the form:
\begin{eqnarray}
  \label{onel}
  \text{in one loop:} &
  \lambda U'(\bar\chi) \bar\Lambda^2,\ 
  \lambda^2(U'(\bar\chi))^2\ln\bar\Lambda ,\\
  \label{twol}
  \text{in two loops:} &
  \lambda U^{(IV)}(\bar\chi) \bar\Lambda^4,\ 
  \lambda^2 (U'')^2 \bar \Lambda^2,\
  \lambda^3 U^{(IV)} (U')^2(\ln\bar\Lambda)^2 ,
\end{eqnarray}
where $\bar\Lambda$ is the regularization parameter (not to be
confused with the tree-level unitarity violation scale of
section~\ref{sec:treeunitarity}).  The first observation is that the
divergences have (generally) a functional form different from the
original potential, so they cannot be reabsorbed in the redefinition
of the coupling constants.  The usual exception to this situation
corresponds to the renormalizable theories, when the potential is a
power series of low enough order.  In a generic case, we obtain an infinite
number of terms without any evident hierarchy between them.

The possibility for the inflationary case under study is to assume
the complete absence of quadratic divergences.  In this case, the higher
order corrections are suppressed by higher powers of $\lambda$.  Note
that the theory still needs further determination.  First, there is
the usual effective theory-like determination---at each order in
$\lambda$ there are divergences from lower order terms, leading to the
counter-terms that should be fixed from measurements.  Second, it is
also possible to add any functional $\chi$ dependence at each order in
$\lambda$ directly to the tree-level potential.  One simple choice to
define a theory is to postulate the tree-level potential (\ref{U}) and
make all calculations in dimensional regularization (thus discarding
automatically all power-law divergences), and choosing a fixed
subtraction scheme (for example, $\overline{\text{MS}}$).  This logic
was implicitly used in all works relating electroweak parameters (the
Higgs boson mass) to inflationary observations
\cite{Bezrukov:2008ej,Barvinsky:2009fy,DeSimone:2008ei}.  Note that
one can change the subtraction rule by an order 1 constant without
significant impact on the results \cite{Bezrukov:2010jz}.

\section{Modification of CMB predictions and bounds on the Higgs boson
  mass}
\label{sec:mass}

Using the described approach, one can calculate the loop-corrected
potential for the inflation.  The recipe described in the end of the
previous section means writing a Coleman--Weinberg effective potential
(for simplicity, we will work exclusively in the Einstein frame in this
section), which has the form $m^4\ln m^2$ for each mass state where
the mass is evaluated in the given background.  The contribution from
the Higgs boson itself is a bit subtle, as its mass squared
$m^2_\chi(\chi)=U'(\chi)$ is negative during inflation; however, this
complication is connected with the fact that we are using the
effective potential calculation away from its minimum, while we should
use the time-dependent approach as Schwinger--Keldysh or `in--in'
formalism \cite{Weinberg:2005vy,George:2012xs}.  However, the Higgs
boson mass at inflation is exponentially small, and the corrections
due to its contribution are suppressed by an extra power of $\xi$.  Thus,
working at the lowest order in $1/\xi$ we can neglect this
contribution.  The contributions of the other SM fields are not small
compared to the tree-level potential
\begin{equation}
  \label{Veff1}
\fl  \Delta U_1 =
  \frac{6m_W^4}{64\pi^2}\left(\ln\frac{m_W^2}{\mu^2}-\frac{5}{6}\right)
  +\frac{3m_Z^4}{64\pi^2}\left(\ln\frac{m_Z^2}{\mu^2}-\frac{5}{6}\right)
  -\frac{3m_t^4}{16\pi^2}\left(\ln\frac{m_t^2}{\mu^2}-\frac{3}{2}\right).
\end{equation}
The $\mu$ dependence here is spurious, and is compensated by the
renormalization group (RG) running of the coupling constants.  With the RG
improvement one can choose $\mu=\mu(\chi)$ in some way to minimize the
contribution of the logarithms.  The result is the (tree-level)
RG-enhanced potential
\begin{equation}
  \label{URGE}
  U_\text{RGE} = \frac{\lambda(\mu(\chi))M_P^4}{4\xi^2(\mu(\chi))}
    \left(1-\e^{-\frac{2\chi}{\sqrt{6}M_P}}\right)^2 .
\end{equation}
One can add the corrections from $\Delta U_1$, associated with the
difference of the particle masses, but they are changing
logarithmically slowly compared to the tree-level contribution, and we
will skip them in the following discussion (in numerical analysis they
are easy to take into account, and do not lead to a significant
modification of the inflationary results).  The RG dependence in the
inflationary region is different from the ordinary SM renormalization
flow and can be obtained exactly from the requirement to cancel the
$\mu$ dependence in (\ref{Veff1}), see \cite{Bezrukov:2009db}.  The RG
equations are known in two cases---either the inflationary case with
$h(\chi)>M_P/\sqrt{\xi}$, or the SM region $H(\chi)<M_P/\xi$ (where it
is the usual SM RG running).  The two RG flows can be connected at any
scale between $M_P/\xi$ and $M_P/\sqrt{\xi}$, without a significant
change of the inflationary potential \cite{Bezrukov:2009db}, which
makes us expect that the effects between these scales (where detailed
analysis is impossible) are weak and can be neglected.

The analysis of the impact on for the inflationary observations was
performed in a set of works
\cite{Bezrukov:2008ej,Bezrukov:2009db,Barvinsky:2009fy,Barvinsky:2008ia,Barvinsky:2009ii,DeSimone:2008ei}.
There is a set of differences in the analysis which deserves a
comment.  Leaving out the earlier works, I will focus on the latest
results in \cite{Bezrukov:2009db,Barvinsky:2009fy,Barvinsky:2009ii}.
There are two significant differences.  First is related directly to
the arbitrariness of the subtraction procedure used to define the
theory, and is inevitable.  Instead of using a constant $\mu$ in the
dimensional regularization in (\ref{Veff1}), it is possible to use
some arbitrary function of the field $\mu=\hat\mu F(\chi)$, where
$\hat \mu$ is used to improve the potential with the usual RG
corresponding to changing momentum, and $F(\chi)$ represents the
arbitrariness in the subtraction procedure\footnote[1]{In the textbook
  formulation of the quantum field theory, the function $F(\chi)$ is
  set to 1, because this is the choice for the renormalizable
  theories.  If the subtraction is being made with other goals in
  mind, for example, to retain the scale invariance of the theory on
  the quantum level, other choices can be made
  \cite{Shaposhnikov:2008xi,Englert:1976ep}.  As far as the
  inflationary theory is non-renormalizable from the start, the choice
  $F=1$ is not obviously the best one.}.  Two choices of the function
$F(\chi)$ are usually considered:

\begin{center}
  \begin{tabular}{lcc}
    \toprule
    $\mu^2/M_P^2\propto$  & Einstein frame & Jordan frame \\
    \midrule
    Choice I
      & $F^2_I = 1$ & $\displaystyle F^2_I = \frac{M_P^2+\xi h^2}{M_P^2}$ \\
    Choice II
      & $\displaystyle F^2_{II} = \frac{M_P^2}{M_P^2+\xi h^2}$ &
      $F^2_{II} = 1$\\
    \bottomrule
  \end{tabular}
\end{center}

\begin{figure}
  \centering
  \includegraphics[width=0.5\textwidth]{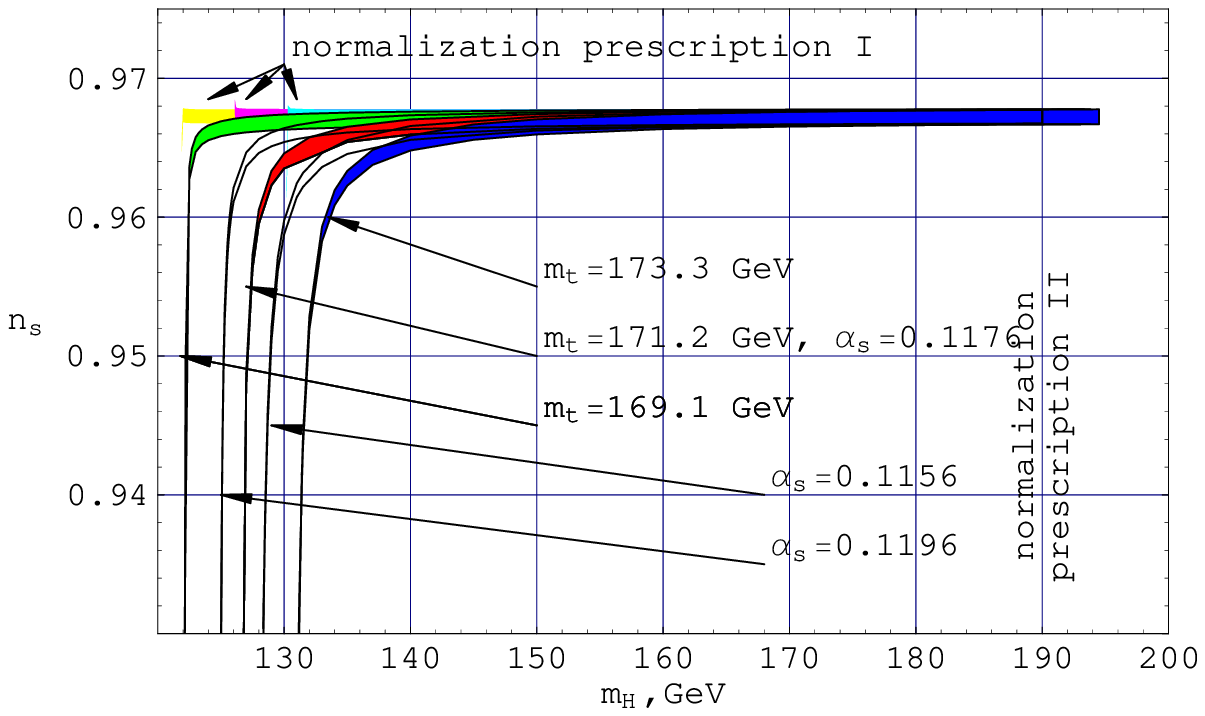}%
  \includegraphics[width=0.5\textwidth]{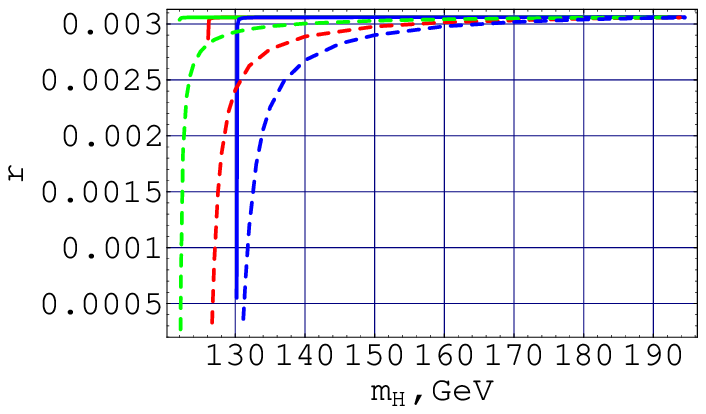}
  \caption{The predictions for the spectral index $n_s$ (left) and
    tensor-to-scalar ratio $r$ (right) with radiative corrections
    \cite{Bezrukov:2009db}.  Note that for the scale-invariant normalization
    prescription I, the predictions coincide with the
    tree level, while in the prescription II, the difference
    becomes more significant at low values of the Higgs mass.}
  \label{fig:nsr_rgh}
\end{figure}

The first one is the usual choice for the Einstein frame.  At the same
time, it is the choice motivated by the scale-invariant quantization
in the Jordan frame, where the scale used for the quantum corrections
is proportional to the effective Planck mass in the Higgs field
background \cite{Bezrukov:2012hx}.  The second choice is usually
explained as the Jordan frame quantization---it corresponds to the
constant scale (ordinary quantization rules, breaking the scale
invariance at a high Higgs field background) in the Jordan frame, and
looks quite peculiar in the Einstein frame.  In the studies
\cite{Bezrukov:2008ej,Bezrukov:2009db}, both choices were analysed,
while studies \cite{DeSimone:2008ei,Barvinsky:2009ii,Barvinsky:2009fy}
use choice II only.  Using
these choices one obtains the RG-enhanced potential (\ref{URGE})
taking
\begin{equation}
  \mu_I^2(\chi) = \frac{y_t^2}{2}
                  \frac{h(\chi)^2}{1+\xi h(\chi)^2/M_P^2} ,\qquad
  \mu_{II}^2(\chi) = \frac{y_t^2}{2} h(\chi)^2.
\end{equation}
In choice I, the scale $\mu(\chi)$ changes in the same way as the
masses of the particles in the limit of high fields (the masses are
just constant at $\chi\to\infty$ in the Jordan frame).  Note that the
coefficients in front of the logarithms in (\ref{Veff1}) are all
proportional to the tree-level potential (\ref{U(chi)}), it becomes
clear that in this choice the overall contribution of the loop
corrections only changes the overall normalization of the potential,
which leads to the change of the value of $\xi$ in (\ref{xi-numeric}),
but does not change the shape of the potential and the predictions for
the spectral index or tensor-to-scalar ratio.  In choice II, the scale
$\mu_{II}(\chi)$ varies significantly during inflation, leading to
some changes, most significant at lower Higgs boson masses, see figure
\ref{fig:nsr_rgh}.  It should be emphasized that choices of the
subtraction procedure different from I and II are possible, and it is
not clear which is the best one, at least without knowledge of the
full UV complete theory, including gravity.

The second difference between \cite{Bezrukov:2009db} and
\cite{Barvinsky:2009ii,Barvinsky:2009fy,Barvinsky:2008ia,Lerner:2011ge}
is in the coefficient in front of the logarithm (or equivalently to
different forms of RG equations at high values of the Higgs field
background).  The difference is connected with the treatment of the
Goldstone modes of the Higgs field.  In \cite{Bezrukov:2009db}, the
calculation is performed in the Einstein frame in the unitary gauge,
where the Goldstone modes are explicitly massless, and their
contribution to the effective potential is zero.  The studies
\cite{Barvinsky:2009ii,Barvinsky:2009fy,Barvinsky:2008ia} perform
calculations in the Jordan frame, and directly diagonalize the
quadratic action for the excitations on top of the background in it,
using the linear representation of the Higgs field.  In this frame,
the question of the masses of the Goldstone modes is more subtle, as
far as the evolution of the background field should be taken into
account.  In short, in the calculation of \cite{Bezrukov:2009db} the
masses of the Goldstone modes are exponentially suppressed (in a way
similar to the self-contribution of the radial mode driving
inflation), and in
\cite{Barvinsky:2009ii,Barvinsky:2009fy,Barvinsky:2008ia}) they have
the mass of the order of gauge and fermion masses.  Taking them into
account modifies the potential and leads (in prescription II) to
different predictions for the cosmological observables, cf
prescription II lines in figure \ref{fig:nsr_rgh} (figure 7 in
\cite{Bezrukov:2009db}) and figure 3 in \cite{Barvinsky:2009ii}.
Note in passing that the prescription I leads to the CMB predictions
coinciding with the tree level (\ref{gen-nsr}).  The role of Goldstone
bosons for the effective potential in the case of an evolving background
was further analysed in the studies \cite{Mooij:2011fi,George:2012xs},
but only in the context of a minimally coupled theory.  However, it
can be expected that the presence of the non-minimal coupling term
suppresses the Goldstone mode contribution and leads to the result of
\cite{Bezrukov:2009db}.

As far as the inflationary potential is modified, the predictions for
the CMB observables are also changed, figure~\ref{fig:nsr_rgh}
\cite{Bezrukov:2009db}.  In the scale-invariant prescription I,
 the arguments of the logarithms in the loop contributions
to the effective potential (\ref{Veff1}) are nearly constant, so the
overall shape of the potential is not changed from the tree level
(\ref{U(chi)}), and the overall normalization is compensated by a
suitable choice of the free parameter $\xi$.  Thus, in this case the
only effect is the lower bound on the Higgs boson mass, corresponding
to $\lambda(\mu)$ becoming negative above some scale completely
preventing inflation.  In the prescription II, the logarithmic terms in
(\ref{Veff1}) are significant, and the CMB predictions are modified,
with large deviation from the tree-level results at low Higgs boson
masses.

The lower bound on the Higgs mass can be intuitively understood from
the following argument.  If Higgs self-coupling $\lambda$ becomes zero
(or negative) at any scale at or after inflation due to RG evolution,
the second minimum of the RG-improved potential (\ref{URGE}) develops
at this scale, in addition to the ordinary minimum at zero
(electroweak scale).  This will stop the inflation or preheating in
this minimum instead of the one we are living in at present.  To a
good precision in the Higgs mass, this bound can be obtained from the
SM RG running of the coupling constants up to the scale $M_P/\xi$.  As
far as at these scales the logarithmic derivative of $\lambda$ is
small, the exact scale which is used ($M_P/\xi$ or $M_P/\sqrt{\xi}$)
has a weak impact on the Higgs mass.

Careful analysis, involving the multi-loop renormalization running of
the coupling constants between the electroweak and inflationary
scales, and multi-loop relation between the $\overline{\text{MS}}$
constants at the electroweak scale and physical observables (pole
masses), was performed in \cite{Bezrukov:2012sa,Degrassi:2012ry} (see
also \cite{Espinosa:2007qp,Bezrukov:2009db,EliasMiro:2011aa} for
previous results).  The results (for the scale-invariant quantization
I) give the lower bound on the Higgs mass
\begin{equation}
  \label{mmin}
  m_h > \left(
    129.5
    + 1.8\frac{M_t-173.2\,\text{GeV}}{0.9\,\text{GeV}}
    - 0.6\frac{\alpha_s-0.1184}{0.0007}
    \pm 2
  \right)\text{GeV} .
\end{equation}
This is within errors of the actually measured Higgs mass,
$m_h\simeq125.5$\,GeV \cite{CMS:2012gu,ATLAS:2012gk}.

Let us note that the bound on the Higgs mass depends on the way the
model is quantized.  For example, scale-invariant quantization
(prescription I) just leads to the bound (\ref{mmin}).  At the same
time, the choice II may lead to a stronger bound, occurring because of
too small predicted $n_s$ (cf figure~\ref{fig:nsr_rgh}).  Strictly
speaking, this bound also relies on the possibility of relating the
low-energy and high-scale physics, which relies on a relatively smooth
transition through the scale $M_P/\xi$.  Note here that to relax a
bound on the Higgs mass in this way, one should have a significant
difference between $\xi$ at high and low scales, with low scale $\xi$
being very large, significantly reducing the region of energies
($\mu<M_P/\xi$), where the SM RG equations can be used.

Further improvement of the experimental situation is extremely
interesting, as it may rule out the plain Higgs inflationary scenario
or indicate the proper choice for the quantization (I versus II).  To
increase the precision of (\ref{mmin}), one has to take into account
higher order loop calculations, and, most importantly, to improve the
experimental determination of the top mass
\cite{Bezrukov:2012sa,Alekhin:2012py}.

Evading the constraint (\ref{mmin}) is possible by adding scalars to
the theory, which provide additional positive contributions to the
beta function of the Higgs self-coupling, thus loosening the bound
\cite{Clark:2009dc,EliasMiro:2012ay}.

\section{Variations of the Higgs inflation}
\label{sec:variants}

\subsection{Non-minimal derivative coupling}

The non-minimal coupling to the Ricci curvature is the lowest order
operator that can be added to the minimally coupled action.  This term
does not lead to additional degrees of freedom in the theory.  It
turns out that it is possible to construct (uniquely) another term
with the same property which is higher order in
derivatives.\footnote[2]{It is possible to add infinitely many terms
  without derivatives, like $h^n R$.  However, their interest for the
  Higgs inflation is limited, as they lead to the Einstein frame
  potential decreasing at high field values, unless matched with the
  Jordan frame scalar potential $V\propto h^{2n}$ \cite{Park:2008hz}.}
Although normally higher order derivative terms lead to new degrees of
freedom, the term
\begin{equation}
  \label{SGermani}
  \delta S_\text{new} = \int \dd^4x\sqrt{-g} \left[
    -\omega^2 G^{\mu\nu}\dm\Phi^\dagger\partial_\nu\Phi
  \right]
\end{equation}
has a specific cancellation and possesses not more than two time
derivatives \cite{Germani:2010gm}.  Here,
$G^{\mu\nu}=R^{\mu\nu}-\frac{R}{2}g^{\mu\nu}$ is the Einstein tensor,
and $\omega$ is the constant of dimensionality $M^{-2}$.  This term
has an effect of adding significant `friction' to the scalar field
background evolution in the expanding background.  Actually, in the FRW
background $G^{00}\sim H^2$.  If $\omega H\gg1$ the term
(\ref{SGermani}) dominates the kinetic term.  By the field
redefinition $\bar\Phi=\Phi/(\omega H)$ during inflation we return the
time derivative to the canonic normalization, which means an effective
reduction of the quartic coupling constant
$\lambda\to\bar\lambda=\lambda/(\omega H)^4$.  This allows for the
slow roll with $\lambda\sim 1$.

The CMB normalization gives \cite{Germani:2010ux} for the scale
$\omega$ in (\ref{SGermani})
\begin{equation}
  \label{eq:2}
  \omega^{-1} = \left(
    (24\pi^2\Delta_{\cal R}^2)^3 \frac{4}{243\lambda N^5}
  \right)^{1/4} M_P \simeq 5\times10^{-8}\lambda^{-1/4}M_P .
\end{equation}
The predictions for the CMB spectrum and tensor-to-scalar ratio at
the tree level are (see figure~\ref{fig:WMAP})
\begin{equation}
  \label{eq:12}
  n_s \simeq 1-\frac{5}{3N} \simeq 0.972 ,\qquad
  r \simeq \frac{8}{3N} \simeq 0.044 .
\end{equation}
The scale (\ref{eq:2}) required by CMB normalization is rather low,
which could be expected as far as the operator (\ref{SGermani}) is of
a higher order than the non-minimal coupling (\ref{dSNM}).  Thus
(\ref{SGermani}) is normally sub-leading for the HI compared to
(\ref{dSNM}).  Note, however, that applied to other inflationary
models, the term in (\ref{SGermani}) can be important.  An example is
the inflation proceeding not along the modulus of the complex field
$\Phi$, but along its angular direction (as in natural inflation with
an axion) \cite{Germani:2010hd,Germani:2011ua}.  In this case, the term
(\ref{SGermani}) is relevant, while the non-minimal coupling
(\ref{dSNM}) only rescales the overall normalization of the Planck
mass without significantly modifying the shape of the potential.

Some analysis of the unitarity violation in this model can be found in
\cite{Atkins:2010yg}.  Note that similar to the ordinary Higgs
inflation, the cut-off scale is background dependent and becomes large
at inflation.

\subsection{UV-completed Higgs inflation}
\label{sec:unitaryHI}

An important example is the perturbative attempt to provide a
UV completion (see section \ref{sec:treeunitarity}) to the Higgs
inflation which was suggested in \cite{Giudice:2010ka}.  This model
provides the perturbative UV completion up to the Planck scale,
however, at the cost of introducing additional parameters and modifying
the connection between the inflationary and low-energy scales.

An additional scalar field $\sigma$ is introduced with the action
\begin{eqnarray}
  \label{SUnitary}
  S_J = \int & \dd^4x\sqrt{-g}\Bigg[
    - \frac{\left(
        M^2+\xi_\sigma\sigma^2+2\xi\Phi^\dagger\Phi
      \right)}{2}R
    + \frac{1}{2}(\dm\sigma)^2 + |D_\mu\Phi|^2 \nonumber \\
  & - \frac{\kappa}{4}\left(
        \sigma^2-\Sigma^2-2\alpha\Phi^\dagger\Phi
      \right)^2
    - \lambda\left( \Phi^\dagger\Phi-\frac{v^2}{2} \right)^2
  \Bigg] .
\end{eqnarray}
Here, the Higgs vacuum expectation value $v$ is assumed to be
completely negligible for all inflationary and gravitational effects,
while the vacuum value of the field $\sigma$ contributes significantly
to the Planck mass $M_P^2=M^2+\xi_\sigma\Sigma^2$.  In the limit $M=0$
the model of the $\sigma$ field behaves as an induced gravity model
\cite{Zee:1978wi,Smolin:1979uz}.  In the limit $\alpha=0$ the $\sigma$
field is coupled with other fields only via gravity (or is mixed with
all non-conformal terms after the conformal transformation to the
Einstein frame), closely related to the model of
\cite{Spokoiny:1984bd} or $R^2$ gravity \cite{Starobinsky:1980te}.
The non-minimal constant for the field $\sigma$ is assumed to dominate
the Higgs non-minimal coupling, $\xi_\sigma\gg\xi$.  In this limit, the
flat inflationary direction in the Einstein frame is aligned mostly
with the $\sigma$ field than with the Higgs field, and the CMB
normalization now determines the constant $\xi_\sigma$:
\begin{equation}
  \xi_\sigma\sqrt{\frac{\lambda+\kappa\alpha^2}{\kappa\alpha}}\simeq
  5\times10^4
\end{equation}
(cf with (\ref{xi-numeric})).  The predictions for the spectral
index and tensor-to-scalar ratio coincide with the predictions of the
Higgs inflation (\ref{ns-r}) in the large $N$ limit (with probably
slightly modified $N$ because of a different preheating mechanism).
However, strictly speaking, the inflation is no longer governed by the
Higgs field, but by the new field $\sigma$ with its own set of
coupling constants $\xi_\sigma$, $\kappa$.  In the argument of
section~\ref{sec:loops}, this UV completion leads to quadratic
divergences from the point of view of the low-energy theory, i.e.\ the
contributions form the mass of the $\sigma$ particle, thus removing
the connection between the low-energy and inflationary regimes.

In the Higgs sector (and its interaction with gravity), the strong
coupling problem does not arise because $\xi$ is now small.  At
the same time, the sector of the field $\sigma$ and gravity is close to
the induced gravity limit of section~\ref{sec:induced-gravity} (i.e.\
$M^2\sim\xi_\sigma\Sigma^2$), where the inflationary potential is
\emph{exactly} (\ref{U(chi)}), and has the cut-off scale $M_P$,
instead of $M_P/\xi_\sigma$.  Note that one can also think of the
model as a perturbative completion of the HI model, where a new state
($\sigma$) appears with the mass equal to the scale of the expected
violation of the tree-level unitarity.

There were other attempts to construct the unitary completion by the
modification of the kinetic interactions, see
\cite{Lerner:2010mq,Lerner:2011it}.

\subsection{Higgs-dilaton theory}

Another variation of the induced gravity regime is used in the
Higgs-dilaton models
\cite{Shaposhnikov:2008xb,Shaposhnikov:2008xi,GarciaBellido:2012zu,GarciaBellido:2011de,Bezrukov:2012hx}.
The model aims to create a completely scale-free theory, so an
additional dilaton is introduced with the scalar part of the action
\begin{equation}
  \label{SDilaton}
\fl  S=\int \dd^4x\sqrt{-g}\Bigg[
    -\frac{\xi_\sigma \sigma^2+\xi h^2}{2}R
    +\frac{(\partial_\mu\sigma)^2}{2}+\frac{(\partial_\mu h)^2}{2}
    -\frac{\lambda}{4}\left(h^2-\frac{\alpha}{\lambda}\sigma^2\right)^2
    -\beta\sigma^4
  \Bigg].
\end{equation}
If $\beta=0$, then the scale invariance can be broken spontaneously in
the flat space, leading to the generation of the planck mass
$M_P^2=\xi_\sigma\sigma^2$.  If one also assumes that the gravity
satisfies the unimodular constraint $\det(g)=1$, the present-day
cosmological constant effectively emerges in the theory.

The model has a rather similar inflationary behaviour, but with
increasing $\xi_\sigma$ the predictions change, with $n_s$ going out
of the $2\sigma$ allowed region for $\xi_\sigma\gtrsim0.007$.  The
relation is present between the spectral index and the equation of
state for the dark energy predicted by the model, which is an
interesting relation between the properties of the early and late
evolution of the Universe.

\subsection{Other variations}


The important class of models is the supergravity models leading to
non-minimally coupled inflation.  Note, that simple inclusion of
non-zero $\xi$ into the supersymmetric model requires special
formulation of the supergravity, as far as the traditional one leads
to $\xi=0$.  This modification was done in
\cite{Arai:2011aa,Arai:2011nq,Ferrara:2010yw,Kallosh:2010ug,Ferrara:2010in}.
Note, that requirements of the stability of the inflationary potential
mean that the low energy model can not be the simplest MSSM one, but
is an extended NMSSM.  The developed superconformal variation of
supergravity allows for other non-minimally coupled inflationary
models \cite{Kallosh:2013pby}, leading to an interesting class of
supersymmetric inflationary models.  Higgs inflation with additional
scalars can lead to models with a scalar cold dark matter
\cite{Lerner:2009xg,Clark:2009dc,Okada:2010jd,Lerner:2011ge,Gong:2012ri}.
Note also, that if arbitrary modifications of the kinetic term are
allowed, more models leading, at least in the classical level, to the
inflationary behaviour, e.g.\
\cite{Lerner:2010mq,Kamada:2010qe,Kamada:2012se,Nakayama:2010sk}

\section{Conclusions}

The Higgs inflationary model provides a nice and economical setup for
the inflationary expansion of the Universe, without the introduction
of any new particles beyond the already discovered SM ones (one still
needs some new physics, for example, relatively light right-handed
neutrinos, to explain neutrino oscillations, dark matter and baryon
asymmetry of the Universe).  The predictions of the model for the CMB
are in excellent agreement with the latest observations.  However, the
theory is not UV complete, and although all the relevant calculations
in the model are always performed at the momenta below the
(background-dependent) scale, corresponding to the tree-level
unitarity violation, taking into account the quantum corrections
depends on the properties of the (yet unknown) UV completion, and may
modify the model predictions.  Scale invariance is the generic
requirement for the UV completion.

The most interesting future experimental observations, that are
important for the validity of the model, are the polarization of the
CMB and the masses of the Higgs boson and top quark and the SM Higgs
boson properties.  Large $B$-mode polarization of the CMB, meaning a large
tensor-to-scalar ratio, would mean a contradiction with the predictions
of HI.  Further improvement of the Higgs mass measurement together
with the improvement on the top quark mass (possible, for example, on a
lepton collider with the centre of mass energy above 350\,GeV) will
show if the SM vacuum is metastable, or can be stable and allow for
the Higgs inflation.


\expandafter\ifx\csname href\endcsname\relax
  \relax
\else
  \newcommand{\eprint}[2][]{\href{http://arxiv.org/abs/#2}{#2}}
\fi
\bibliographystyle{iopart-num-arxiv}
\begingroup
\footnotesize
\bibliography{HIReview}
\endgroup

\end{document}